\documentclass[conference,onecolumn]{IEEEtran}
\IEEEoverridecommandlockouts

\usepackage{tikz}
\newcommand{\htex}{2in}
\usepackage{subfigure}

\definecolor{Liberty}{HTML}{4357AD}
\definecolor{Burlywood}{HTML}{48A9A6}
\definecolor{FuzzyWuzzy}{HTML}{C1666B}

\definecolor{mDarkBrown}{HTML}{C1666B}
\colorlet{mMediumBrown}{mDarkBrown!80}
\colorlet{mLightBrown}{mDarkBrown!50}

\definecolor{mDarkTeal}{HTML}{4357AD} 
\colorlet{mMediumTeal}{mDarkTeal!80}
\colorlet{mLightTeal}{mDarkTeal!50}

\colorlet{redn}{mDarkBrown}
\colorlet{bluen}{mDarkTeal}

\definecolor{Verdigris}{HTML}{48A9A6}

\usepackage{cite}
\usepackage{amsmath,amssymb,amsfonts}
\usepackage{algorithmic}
\usepackage{graphicx}
\usepackage{textcomp}
\usepackage{xcolor}
\usepackage[ruled,vlined]{algorithm2e}

\usepackage{pgfplots}
\pgfplotsset{compat=newest}
\usepgfplotslibrary{groupplots}

\usetikzlibrary{through,calc}

\def\BibTeX{{\rm B\kern-.05em{\sc i\kern-.025em b}\kern-.08em
    T\kern-.1667em\lower.7ex\hbox{E}\kern-.125emX}}
    
\newcommand{\cel}{{\sf c}}

\newcommand{\bs}{\boldsymbol}
\newcommand{\bb}{\mathbb}
\newcommand{\cl}{\mathcal}

\newcommand{\ts}{\textstyle}
\newcommand{\ie}{\emph{i.e.}, }
\newcommand{\eg}{\emph{e.g.}, }

\newcommand{\tiid}{%
    \ifmmode
        \mathrm{iid}%
    \else%
        iid\xspace%
    \fi%
}

\newcommand{\ong}{\emph{On-the-Grid} }
\newcommand{\ofg}{\emph{Off-the-Grid} }

\newcommand{\sps}[3]{#3\langle#1,\,#2 #3\rangle}

\newcommand{\amin}[1]{\underset{#1}{\text{arg}\,\min}}
    
\begin{document}

\pgfplotsset{every tick label/.append style={font=\scriptsize}}

\title{\textit{Going Below and Beyond}\\
{Off-the-Grid Velocity Estimation\\ from 1-bit Radar Measurements}
}
\author{\IEEEauthorblockN{Gilles Monnoyer de Galland$^\star$, Thomas Feuillen$^\star$, Luc Vandendorpe$^\star$, Laurent
Jacques$^\dagger$\thanks{GM and LJ are funded by the Belgian FNRS.}}
\IEEEauthorblockA{\textit{$^\star$ELEN. $^\dagger$INMA.  ICTEAM, UCLouvain} \\
Louvain-la-Neuve, Belgium \\
\textit{firstname.lastname@uclouvain.be}}}

\maketitle

\begin{abstract}
In this paper we propose to bridge the gap between using extremely low resolution 1-bit measurements and estimating targets' parameters, such as their velocities, that exist in a continuum, \ie by performing \textit{Off-the-Grid} estimation. To that end, a Continuous version of Orthogonal Matching Pursuit (COMP) is modified in order to leverage the 1-bit measurements coming from a simple Doppler Radar. Using Monte-Carlo simulations, we show that although the resolution of the acquisition is dramatically reduced, velocity estimation of multiple targets can still be achieved and reaches performances beyond classic \ong methods. Furthermore, we show empirically that adding a random and uniform dithering before the quantization is necessary when estimating more than one target. 
\end{abstract}


\section{Introduction}

In recent years the use of coarse acquisition such as 1-bit quantization has gain traction in radar applications \cite{xi2019gridless,Feuillen20181bitLS,Feuillen2018QuantityOQ,8567873}. Lowering the resolution of the acquisition helps creating more cost or energy efficient systems or lower the data load in applications with multiple sensors working jointly, such as MIMO systems \cite{xi2019gridless,Zhang2020DeepLF}. Lowering the resolution of the acquisition usually comes at a cost in terms of performances. Imposing more assumptions on the signal is one way to remedy this loss. Such assumptions are, for example, that the signal is sparse, or that the targets of interest exist in a discretized domain (\eg a discrete range profile, velocities that lie on a grid,...). In this paper, we challenge this latter assumption by developing a method that can recover from 1-bit dithered measurements the velocities of different targets that belong in a continuum. 
Simply put, we quantized the measurements with a resolution far \textit{below} the classic acquisition while still being able to estimate targets' velocities \textit{beyond} the constraints of the discrete grid.

Coarse acquisition process such as 1-bit quantization, where only the signs of the measurements are recorded from analog signals, have been used in various applications \cite{xi2019gridless,Feuillen20181bitLS,Feuillen2018QuantityOQ,8682059,xu2018quantized,Kamilov2012OneBitMW,8567873,Stein2016DOAPE}. Given the important data compression that they operate on the sampled signal compared to a high resolution signal, they were used in \cite{8567873} for Synthetic Aperture Radar (SAR) processing. Moreover, with the advent of compressive sensing \cite{Dirksen2019,intro2CS,Foucart2016FlavorsOC,bara}, there has been numerous studies of how to leverage this quantization beyond just treating it as a bounded quantization noise applied on the classic measurements. 
In \cite{xu2018quantized} it was shown that Projected Back-Projection (PBP), which projects the matched filter estimation onto the set of sparse signals, when used in conjunction with a uniform dithering, provides a uniform reconstruction guarantee on sparse signals. This work paved the way to more applied papers on the use of quantized and dithered measurement in radar signal processing \cite{Feuillen20181bitLS,Feuillen2018QuantityOQ}. Other methods rely on the Deep-Learning framework to perform reconstruction of signals \cite{Zhang2020DeepLF}. Finally, there exists other works that rely on more applications based properties of the signals \cite{ Foucart2016FlavorsOC,10.1117/12.2024412,Huynh2018FastBE}.  

On another flavor, the field dedicated to the estimation of signals parameterized over a continuous domain have recently grown in interest.  
For instance, in~\cite{traonmilin2019, keviren2017}, the authors estimate multiple frequencies on a continuous domain from a linear combination of complex exponentials by adaptating a projected gradient descent.
In~\cite{tang2013, mishra2014}, the authors solve the \ofg problem with an Atomic Norm Minimization (ANM) formulated as a semidefinite program.
An application of this strategy for 1-bit quantized radar measurements has been proposed in~\cite{xi2019gridless}.
Although these methods exhibit promising performance, their high computational complexities make them not suitable for real-time radar applications.

In~\cite{simoncelli2011}, a continuous version of the Basis Pursuit~\cite{BP} is derived from the definition of an interpolation model.
This interpolation efficiently approximates any received waveform parameterized over a continuum from a finite set of interpolants. 
In~\cite{knudson2014, duarte2014}, the authors similarly designed a continuous adaptation of the Orthogonal Matching Pursuit (OMP)~\cite{OMP}, namely the Continuous OMP (COMP), using the same interpolation concept. 
The low complexity of this algorithm makes it appealing for real-time radar applications. 
In this paper, we extend COMP to work with 1-bit quantized measurements and evaluate the effect of dithering on this algorithm. 
While other contributions have investigated the combination of quantization and \ofg estimation via ANM~\cite{feng2018,haoyu2018,xi2019gridless}, our implementation of COMP proposes an alternative that is affordable in terms of computation time.

This paper is organized as follow. 
In Section \ref{sec:model}, we present the ideal expression of the signal received by a simple monotone radar in the presence of multiple targets with different velocities. 
Section \ref{sec:quantize} describes the effect of quantizing this signal in the extreme case of 1-bit resolution. 
The use of dithering is motivated and the expression of the resulting signal is detailed.
In Section \ref{sec:COMP}, we present in details our extension of COMP that we use to recover the \ofg values of the velocities from 1-bit quantized signal.
Finally, Section \ref{sec:result} proposes an analysis of the results of extensive Monte-Carlo simulations before concluding in Section \ref{sec:concl}.

The claims of the paper are as follows:
(i) the Monte-Carlo simulations show the ability of our algorithm to recover the targets' velocities from 1-bit quantized measurements by leveraging the effect of the dithering, 
(ii) we show that the improvement brought by the interpolation methodology of COMP holds for quantized measurement and 
(iii) the low complexity of our adaptation of COMP together with 1-bit and dithered measurements may open the way for new designs of cost-effective and high resolution radar systems.

\paragraph*{Notations} Matrices and vectors are denoted by bold symbols, $\mathsf{i}=\sqrt{-1}$, and $\cel$ is the speed of light. The scalar product between the vectors $\bs{a}$ and $\bs{b}$ reads $\sps{\bs{a}}{\bs{b}}{}$. 
The transpose and conjugate transpose of a matrix $\bs{A}$ are $\bs{A}^\top$ and $\bs{A}^H$, 
$\bb C\cl N(0,\sigma^2)$ denotes the complex normal distribution with variance $\sigma^2$, $\cl U(a,b)$ denotes the uniform distribution with $a<b$, 
$\mathrm{d_T}(a,b):=\min(|b-a|, |1-(b-a)|)$ is the unit 1D torus distance between $a$ and $b$. we write $[K] := \{1,\cdots,K\}$.

\section{System Model}
\label{sec:model}
In this paper, we study a simple Continous Wave Radar (CWR) transmitting a single frequency $f_0$.
The observed scene is assumed to contain a known number $K$ of distinct targets moving with the respective radial velocities $\{v_k\}_{k=1}^K$. 
Assuming a noiseless scenario, after a coherent demodulation of the signal resulting from the $K$ received echoes and a sampling at rate $1/T_s$, we obtain the signal
\begin{equation}
    \ts \bs y = \sum_{k=1}^K \alpha_k \bs a(v_k),
    \label{eq:y}
\end{equation}
where $\bs a(v) := (a_1(v), \cdots, a_M(v))^\top \in \bb C^M$ and
\begin{equation}
     a_m(v) = \exp(-\mathsf{i} 4\pi f_0 T_s \frac{v_k}{\cel} m).
     \label{eq:a}
\end{equation}
The coefficients $\alpha_k\in \bb C$ are independant random variables that model all attenuation and phase shift occurring in the transmission.  
To meet the Shannon-Nyquist sampling theorem, the velocities are assumed to be bounded such that
$v_k \in \cl V := [-\frac{\cel}{4 f_0 T_s}, \frac{\cel}{4 f_0 T_s}[$ for all $k\in[K]$.

The equation \eqref{eq:y} represents the signal $\bs y$ with a linear combination of $K$ \emph{atoms} $\bs a(v)$ parameterized in $\cl V$.
In other terms, $\bs y$ is decomposed in the continuous set of waveforms, named a dictionary, $\cl D := \{\bs a(v)\}_{v\in\cl V}$.
Such a general representation was considered earlier in continuous sparse signal recovery~\cite{gribonval2001}.


\section{Quantization}
\label{sec:quantize}
A full resolution model of the CWR signal has been introduced in the previous section. For this signal to be processed it first needs to be measured. The digitisation of analog signals is performed using Analog to Digital Converters (ADCs). Their fixed resolution induces a discrepancy between the analog signal and its measured version. Classically, for high resolutions, this discrepancy is modelled as a random and bounded quantization noise or is simply ignored in the processing. The latter can only be done when the ADCs' resolution is high enough. In this work we decide to dramatically depart from this requirement by reducing the resolution of the ADCs to one bit; effectively only recording the $sign$ of the measurements on each \textit{I} and \textit{Q} channels after the coherent demodulation. 

We define the quantization operator as $$\cl{Q}_\delta(\bs y)= \frac{\delta}{2} \text{sign}(\bs y^{\mathbb{R}})+ \mathsf{i} \frac{\delta}{2} \text{sign}(\bs y^{\mathbb{I}}), $$

with $\delta \geq 2 \max_m(|y_m|)$, and $\bs y = \bs y^{\mathbb{R}}+ \mathsf{i} \bs y^{\mathbb{I}}$.

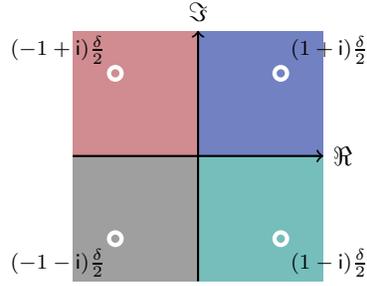
\begin{figure}[h]
    \centering
       \usetikzlibrary{through,calc}

\usetikzlibrary{shadings}

\definecolor{bblue}{HTML}{4F81BD}
\definecolor{rred}{HTML}{C0504D}
\definecolor{ggreen}{HTML}{9BBB59}
\definecolor{ppurple}{HTML}{9F4C7C}
%
%
%
%
%


\def\axesradargrid at (#1,#2){\draw[help lines,step=0.5] (0,0) grid (#1,#2);
\draw[->, thick](0,#2/2)--(#1,#2/2);
\node [right] at (#1,#2/2) {$ \Re$};
\draw[->,thick](#1/2,0)--(#1/2,#2);
\node [above] at (#1/2,#2) {$\Im$};}

\def\axesradar at (#1,#2){
\draw[->, thick](0,#2/2)--(#1,#2/2);
\node [right] at (#1,#2/2) {$ \Re$};
\draw[->, thick](#1/2,0)--(#1/2,#2);
\node [above] at (#1/2,#2) {$\Im$};}

\def\pointS at (#1:#2){
\draw [mDarkTeal, thick,fill=mMediumTeal] (#1:#2) circle [radius=0.04];
\draw [mMediumBrown, thick,dashed] (#1:#2) circle [radius=0.2];}

\begin{tikzpicture}[thick,scale=5/3, every node/.style={scale=3/3}]


\begin{scope}[shift={(1,1)}]

%
%
%
%
%
%
%
%
%
%
%
%
 

\fill [mDarkTeal, opacity=0.75] (0,0) rectangle (1,1);
 
 \fill [mDarkBrown, opacity=0.75] (0,0) rectangle (-1,1); 
  
 \fill [Burlywood, opacity=0.75] (0,0) rectangle (1,-1); 


\fill [gray, opacity=0.75] (0,0) rectangle (-1,-1);    

\draw [white,ultra thick] (0.66,0.66) circle [radius=0.05];
\node[above right] at (0.66,0.66)  {\footnotesize $(1+\mathsf{i})\frac{\delta}{2}$};

\draw [white,ultra thick] (0.66,-0.66) circle [radius=0.05];
\node[below right] at (0.66,-0.66)  {\footnotesize $(1-\mathsf{i})\frac{\delta}{2}$};

\draw [white,ultra thick] (-0.66,-0.66) circle [radius=0.05];
\node[below left] at (-0.66,-0.66)  {\footnotesize $(-1-\mathsf{i})\frac{\delta}{2}$};

\draw [white,ultra thick] (-0.66,0.66) circle [radius=0.05];
\node[above left] at (-0.66,0.66)  {\footnotesize $(-1+\mathsf{i})\frac{\delta}{2}$};

\end{scope}

\axesradar at (2,2)
\end{tikzpicture}
    \caption{Representation of the 1-bit quantization of resolution $\delta$ performed on both the real and imaginary domain.}
    \label{fig:quadrant}
\end{figure}

However, authors in \cite{Plan2014} have warned against using this quantization directly on signal made of a linear combination of few components, which is are complex exponentials in our case. Indeed, Plan and Vershynin showed, for instance, that binary waveforms (\eg Bernoulli)
have intrinsic limitations. For example, some weights $\alpha_k$ in \eqref{eq:y} cannot be recovered  from their quantized measurements $\mathcal{Q}_\delta (\bs y) $ regardless of the number of measurements $M$.
More recently, authors in \cite{Feuillen20181bitLS,Feuillen2018QuantityOQ} showed a similar issue for Fourier based model. They showed theoretically and through radar measurements that information related to low power targets can be entirely removed from the quantized measurements, thus preventing any further signal estimation.
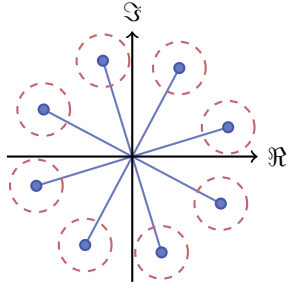
\begin{figure}[h]
    \centering
 \usetikzlibrary{through,calc}

\usetikzlibrary{shadings}

\definecolor{bblue}{HTML}{4F81BD}
\definecolor{rred}{HTML}{C0504D}
\definecolor{ggreen}{HTML}{9BBB59}
\definecolor{ppurple}{HTML}{9F4C7C}

%
%
%


\def\axesradargrid at (#1,#2){\draw[help lines,step=0.5] (0,0) grid (#1,#2);
\draw[->, thick](0,#2/2)--(#1,#2/2);
\node [right] at (#1,#2/2) {$ \Re$};
\draw[->,thick](#1/2,0)--(#1/2,#2);
\node [above] at (#1/2,#2) {$\Im$};}

\def\axesradar at (#1,#2){
\draw[->, thick](0,#2/2)--(#1,#2/2);
\node [right] at (#1,#2/2) {$ \Re$};
\draw[->, thick](#1/2,0)--(#1/2,#2);
\node [above] at (#1/2,#2) {$\Im$};}

\def\pointS at (#1:#2){
\draw [mDarkTeal, thick,fill=mMediumTeal] (#1:#2) circle [radius=0.04];
\draw [mDarkBrown, thick,dashed] (#1:#2) circle [radius=0.2];}

\def\pointC at (#1:#2:#3:#4){
\draw[#3, thick] (0,0)--(#1:#2) ;
\draw [#4, thick,fill=#3] (#1:#2) circle [radius=0.04];}

\begin{tikzpicture}[thick,scale=5/3, every node/.style={scale=3/3}]


\begin{scope}[shift={(1,1)}]

\draw[help lines,step=0.25,white,opacity=0.0] (-1.25,-1.25) grid (1.25,1.25);

\def\angleSA{17}
\def\angleSB{110}

\def\radiusSB{0.21}

\def\omegaSA{45}

\pointC at (\angleSA:0.8:mMediumTeal:mDarkTeal);

\pointC at (\angleSA+\omegaSA:0.8:mMediumTeal:mDarkTeal);

\pointC at (\angleSA+\omegaSA*2:0.8:mMediumTeal:mDarkTeal);

\pointC at (\angleSA+\omegaSA*3:0.8:mMediumTeal:mDarkTeal);

\pointC at (\angleSA+\omegaSA*4:0.8:mMediumTeal:mDarkTeal);

\pointC at (\angleSA+\omegaSA*5:0.8:mMediumTeal:mDarkTeal);
\pointC at (\angleSA+\omegaSA*6:0.8:mMediumTeal:mDarkTeal);
\pointC at (\angleSA+\omegaSA*7:0.8:mMediumTeal:mDarkTeal);

\draw [mDarkBrown, thick,dashed] (\angleSA+\omegaSA*4:0.8) circle [radius=\radiusSB];
\draw [mDarkBrown, thick,dashed] (\angleSA+\omegaSA*3:0.8) circle [radius=\radiusSB];
\draw [mDarkBrown, thick,dashed] (\angleSA+\omegaSA*2:0.8) circle [radius=\radiusSB];
\draw [mDarkBrown, thick,dashed] (\angleSA+\omegaSA*1:0.8) circle [radius=\radiusSB];
\draw [mDarkBrown, thick,dashed] (\angleSA+\omegaSA*5:0.8) circle [radius=\radiusSB];
\draw [mDarkBrown, thick,dashed] (\angleSA+\omegaSA*6:0.8) circle [radius=\radiusSB];
\draw [mDarkBrown, thick,dashed] (\angleSA+\omegaSA*7:0.8) circle [radius=\radiusSB];
\draw [mDarkBrown, thick,dashed] (\angleSA+\omegaSA*0:0.8) circle [radius=\radiusSB];

\end{scope}

\axesradar at (2,2)
\end{tikzpicture}
    \caption{Representation of an artefact where two different full resolution signals (in \textcolor{redn}{red} and \textcolor{bluen}{blue}), have the same 1-bit measurements.}
    \label{fig:artefact}
\end{figure}

Fig. \ref{fig:artefact} is the representation of an artefact where a signal resulting from 1 target -- whose corresponding measurements are represented by the blue points --  and from two targets -- represented by the red circles -- are, once quantized, sent to the exact same quadrants removing any possibility of a successful estimation.

This issue, coming from the deterministic way in which the quantizer operates, can be solved by dithering the signal. In other words, by changing the thresholds used in the quantization at each measurement. The quantizer then becomes :
\begin{equation}
\cl A_\delta (\bs y ) = \cl Q_\delta (\bs y + \bs \xi)
\end{equation}
where $\bs \xi \in \mathbb{C}^{M}$ is the dithering. There exists different strategies that one can follow to set this dithering \cite{9239262,Kamilov2012OneBitMW, xu2018quantized,Dirksen2019}. In this paper we focus on the use of a random uniform dithering that spans the dynamic of the measured signal. While adding this random vector seems counter intuitive at first for enhancing the estimation, authors in \cite{xu2018quantized} showed that adding a dithering that follows a uniform distribution $\xi_i \sim \mathcal{U}(-\frac{\delta}{2}, -\frac{\delta}{2})+ \mathsf{i}\mathcal{U}(-\frac{\delta}{2}, -\frac{\delta}{2})$ allows to upperbound the reconstruction of sparse \ong signals using the Projected Back-Projection (PBP) algorithm. This bound relies on the fact that for a uniform dithering with an appropriate $\delta$, $\mathbb{E}(\cl A_\delta (\bs y ) )= \bs y$. 
The upperbound on the signal reconstruction of PBP scales as $\mathcal{O}(\sqrt{\frac{K}{M}})$. This means that, in order to obtain a better estimate, and thus avoid these artefacts, one only need to increase the number the measurements $M$. 

\section{Off the Grid Reconstruction}
\label{sec:COMP}
The velocities of the $K$ targets observed by the radar are taken from the continuous space $\cl V$.
Conventionally, these velocities are assumed to lie on a grid which results from the discretization of $\cl V$.
In this paper, we aim to break this common assumption and compute \ofg estimates of the velocities. 
In this framework, we consider the Continuous Orthogonal Matching Pursuit (COMP) algorithm which has been developed in \cite{knudson2014, duarte2014} and extended in~\cite{monnoyer2020} for radar applications. 
The COMP extends the Orthogonal Matching Pursuit (OMP)~\cite{OMP} with the concept of interpolant dictionaries from the continuous Basis Pursuit~\cite{simoncelli2011, duarte2013}. 
This greedy algorithm has the advantage to be sufficiently fast for real time radar applications~\cite{monnoyer2020} while being naturally compatible with the setting of 1-bit measurements presented in the previous section.

In the next paragraphs, we briefly describe the general principle of OMP and COMP and explain how COMP succeeds to provide \ofg estimates. 
Then, we describe how we adapted COMP to quantized measurements.

\subsection{From OMP to COMP}

Both OMP and COMP operate with a parameter grid which results from the discretization of $\cl V$. 
Let us denote this grid by $\Omega := \{\bar{v}_n\}_{n=1}^N\subset\cl V$.
In OMP, we build a dictionary which contains the $N$ atoms associated to $\Omega$, \ie this dictonary is $\cl D_{\Omega} := \{\bs a(v)\}_{v\in\Omega}$. Then, in a nutshell, OMP attempts to greedily determine the best linear combinations of $K$ distinct atoms taken from $\cl D_{\Omega}$ to fit the measurements.
Therefore, although the actual velocities of the targets may take values outside of $\Omega$ (\ie \ofg) OMP is only able to provide \ong estimates, leading to grid errors.
In COMP, we instead build a group of interpolant dictionaries which enables us to approximately access, from the grid, all the atoms that lie \ofg while using finite dictionaries.
In other terms, we associate to each grid index $n\in[N]$ a group of $I$ distinct interpolant atoms denoted by $\{\bs d_i[n]\}_{i=1}^I$.
With a sufficiently dense grid, we assume that for all $v\in\cl V$, there is a grid bin $\bar v_n\in\Omega$ indexed by $n$ such that
\begin{equation}
    \ts \bs a(v) \simeq \sum_{i=1}^I \bs d_i[n] c^i_n(v),
    \label{eq:interp_scheme}
\end{equation}
where the coefficients $c^i_n(v)$ are defined from a mapping function~\cite{knudson2014} $\cl C(v - \bar{v}_n)$ which depends on the interpolation scheme, \ie $\bs c_n(v) := (c^1_n(v), \cdots, c^I_n(v))^\top = \cl C(v - \bar{v}_n)$.
For example, if one builds an interpolation scheme from an order-2 Taylor expansion, then $I=3$ and for $i=1..3$ the interpolant $\bs d_i[n]$ is the $(i-1)$-th derivative of $\bs a(v)$ evaluated in $\bar v_n$ and $\cl C(v-\bar v_n)=(1, v-\bar v_n, \frac{(v-\bar v_n)^2}{2})^\top$. 

We denote $\bs D_n := [\bs d_1[n], \dots, \bs d_I[n]]$ for the simplicity of notations.
With the approximation \eqref{eq:interp_scheme}, we can write that there exist at least one of set grid bins indexed by $n_1, \cdots, n_k$ such that
\begin{equation}
    \ts \bs y \simeq \sum_{k=1}^K \bs D_{n_k} \bs \beta_k,
\end{equation}
where $\bs \beta_k := \alpha_k\bs c_{n_k}(v) \in \bb C^I$.


\begin{algorithm}[tb!]
\SetKwInOut{Input}{Input}\SetKwInOut{Output}{Output}
\SetKwFor{While}{While}{:}{end}

\Input{$K$, $\bs z$, $\bs \xi$}

\Output{$\{\hat{\alpha}_k\}_{k=1}^K, \{\hat v_k\}_{k=1}^K$} 

\Begin{

Initialization: $\bs{r}^{(1)} = \bs{z}$, $\bar{\bs D}^{(0)} = \emptyset$, $k=1$;
\vspace{2mm}

\While{$k \leq K$}{
    
    \begin{equation}
    \!\!\!\!\!\!\!\!\!\!\!\!\!\!\!\!\!\!\!\!\!\!\!\!\!\!\!\!\!\!\!\!\!\!\!\!\!\!\!\!\hat{n}_k = \amin{n\in[N]}\big|\sps{\bs a(\bar{v}_n)}{\bs r^{(k)}}{}\big|
    \label{eq:index}
    \end{equation}
    \vspace{-4mm}
    
    \begin{flalign}
    &\bar{\bs{D}}^{(k)} = \big[\bar{\bs{D}}^{(k-1)}, \bs D_{\hat{n}_k}\big]&
    \label{eq:Dbar}
    \end{flalign}
    \vspace{-4mm}
    
    \begin{flalign}
    &\hat{\bs \beta}^{(k)} = \amin{\bs \beta\in\bb C^{kI}}\|\bar{\bs D}^{(k)}\bs \beta - \bs z\|_2^2&
    \label{eq:beta}
    \end{flalign}
    \vspace{-4mm}
    
    \begin{flalign}
    &\bs r^{(k+1)} = \bs z - \cl A\big(\bar{\bs D}^{(k)}\hat{\bs \beta}^{(k)}\big)&
    \label{eq:rem_contr}
    \end{flalign}
    
     $k \leftarrow k+1$
}
\vspace{2mm}
    \textbf{For all} $k\in[K]$,\\
    \vspace{-3mm}
    
    \begin{equation}
        (\hat{\alpha}_k, \hat{v}_k)^\top = \amin{\alpha\in\bb C, v\in\cl V} \big\|\hat{\bs \beta}_k -  \alpha \cl C(v - \bar{v}_{\hat{n}_k}) \big\|_2^2
        \label{eq:corr_step}
    \end{equation}
    \vspace{-2mm}
}
  \caption{Quantized COMP (QCOMP)}
  \label{alg:COMP}
\end{algorithm}

\subsection{COMP for quantized radar measurements}
Alg. \ref{alg:COMP} directly formulates the Continuous OMP for quantized measurement. We named this adaptation the Quantized Continuous OMP (QCOMP).
In each iteration of this greedy algorithm, QCOMP first determines from \eqref{eq:index} the index of the \ong atom that best fits the residue $\bs r^{(k)}$.
This residue is initially set to the measurement $\bs z$. 
Such correlation, similar to PBP, has been shown to be efficient even with the 1-bit quantized measurement $\bs z := \cl A(\bs y)$. 
As stated in the previous section, dithering the measurements is required to enable the successful estimation of more than one target.
We show in the next section the relevance of this result for OMP-based algorithms.

In the next step, we estimate the coefficients $\bs \beta^{(k)} := (\bs \beta_1^\top, \cdots \bs \beta_k^\top)^\top$.
More precisely, in each iteration, we select the coefficients that provide the linear combination of $k\times I$ interpolant atoms from $\bar{\bs D}^{(k)}$ that best fit $\bs z$.
As QCOMP works with the quantized measurement $\bs z:= \cl A(\bs y)$, one may want to estimate these coefficients with the minimization
\begin{equation}
   \ts \hat{\bs \beta}^{(k)} =  \amin{\bs \beta\in\bb C^{kI}}\|\cl A(\bar{\bs D}^{(k)}\bs \beta) - \bs z\|_2^2.
   \label{eq:betabetter}
\end{equation}
Solving \eqref{eq:betabetter} however is challenging and would tremendously increase the computation time of QCOMP with respect to COMP.
Therefore, we chose to perform the minimization of \eqref{eq:beta} which is quickly solved with the computation of a pseudo-inverse.
In the future, we may investigate fast heuristic method to solve \eqref{eq:betabetter}.

At the end of each iteration, the residue is updated by removing from $\bs z$ the \emph{quantized and dithered} $k$ atoms estimated from the $k\times I$ interpolants.

Finally, when $I>1$ the definition of the mapping function $\cl C(v-\bar v_n)$ implies that $\hat{\bs \beta}$ contains information about both $\alpha_k$ and the \ofg deviation $v_k-\bar v_{\hat n_k}$.
This information is recovered in the final step defined by eq \eqref{eq:corr_step} by projecting $\hat{\bs \beta}_k$ on the feasible set of $\bs \beta_k$ separately for each $k\in[K]$. 
The separation of steps \eqref{eq:beta} and \eqref{eq:corr_step} is essential to guarantee a time-efficient algorithm and is only poorly detrimental to the reconstruction~\cite{knudson2014}.



\section{Simulations results}
\label{sec:result}
We evaluated the ability of QCOMP to recover the velocities of $K$ distinct targets with extensive Monte-Carlo simulations.
We compared two different reconstruction schemes: (S1) a non-interpolated scheme with $I=1$ and $\bs d^{(1)}[n] := \bs a(\bar{v}_n)$ and (S2) the interpolation scheme built from an order 1 Taylor expansion of the grid atoms, \ie with $I=2$, $\bs d^{(1)}[n] = \bs a(\bar{v}_n)$, $\bs d^{(2)}[n] = \frac{\partial\bs a}{\partial v} (\bar{v}_n)$~\cite{simoncelli2011, knudson2014}.  
Note that COMP with the scheme (S1) is strictly equivalent to OMP and hence, is not able to provide \ofg estimates. This will act as a reference to highlight the benefit of using interpolant atoms (even with a simple order 1 Taylor expansion) in (S2).

For each reconstruction scheme, we consider three different measurement models from which the reconstruction operates: (a) the classic measurement $\bs y$, (b) the 1-bit quantized measurement $\bs z$ with no dithering ($\bs \xi = 0$) and (c) the quantized dithered signal with $\bs \xi$ following the complex uniform distribution presented in Section \ref{sec:quantize}. 

\subsection{Performance metrics and settings}
In Fig. \ref{fig:K1} and \ref{fig:K2}, we compare the performance of these six different scenarios.
Each point in each curve is obtained from the average results of 10,000 independent realisations of $K$ targets from the distributions $\alpha_k\sim\bb C\cl N(0,1)$ and $v_k\sim \cl U(\cl V)$. 
We defined four performance metrics which together provide a complete and thorough evaluation of the different scenario. The first three are related to the support recovery capability of the algorithm. 
Each run of the algorithm in a given scenario provides $K$ velocity estimates. With an appropriate choice of association between the estimates $\{\hat v\}_{k=1}^K$ to the actual values $\{v\}_{k=1}^K$, we define the estimation error for all $k\in[K]$ as $E_k := \mathrm{d_T}(\hat v_k, v_k) / (\frac{\cel}{2f_0T_sM})$. 
The factor $\frac{\cel}{2f_0T_sM}$ corresponds to the intrinsic resolution (\ie the width of the main lobe) of the ambiguity function.
This normalization enables our results to be general for any radar system parameter values as the velocity estimation error is computed with respect to this resolution.

The four performance metrics are defined as follows.
\begin{itemize}
    \item The average estimation defined by the average value of $K^{-1}\sum_{k=1}^KE_k$ across the 10,000 realisations,
    \item The miss rate: the estimation of $v_k$ is a \emph{miss} if $E_k\geq1$,
    \item The average hit error that computes the estimation error only across the unmissed estimations,
    \item The average residue or reconstruction error defined by $\|\bs y-\bar{\bs D}^{(K)}\hat{\bs \beta}^{(K)}\|_2$ when $\bs y$ is used as input for COMP and $\|\bs z-\cl A(\bar{\bs D}^{(K)}\hat{\bs \beta}^{(K)})\|_2$ when $\bs z$ is the input of QCOMP.
\end{itemize}

For each targets' realisation, the estimates $\{\hat v_k\}_{k=1}^K$ are paired with the actual velocities in order to have the lowest miss rate.

The result of the next section are obtained with $M=256$ and by modifying the grid density which we define as the 
ratio $N/M$.
Increasing this density improves the ability of grid based algorithms ---such as QCOMP--- to estimate parameter values that lie on a continuous domain~\cite{simoncelli2011, knudson2014, monnoyer2020}.
The non interpolated scheme cannot provide estimates lying outside the grid and hence, theoretically requires an infinite $N$ to possibly estimate the exact velocities. 
The interpolation scheme enables us to estimate \ofg velocities from a grid with reasonable density. 
Therefore with classical measurements, increasing $N/M$ continuously improves the overall quality of estimation at the cost of computation time. 
Moreover for a fixed density $N/M$, an interpolated model provides better estimates than a non interpolated one~\cite{simoncelli2011, monnoyer2020}.
In the next section, we study how this statement is affected by the quantization either with and without dithering.



\subsection{Results}


Now that the simulation's setting and metric have been introduced, let us focus on the results. 
As expected, in Fig \ref{fig:K1}(d), we see that the classic acquisition without quantization exhibits the best reconstruction performances. Furthermore, one can observe that using the interpolant dictionaries based on the Taylor approximation allows for a finer estimation of the velocity of the target, which in turns helps reducing the residue. The miss rate represented in Fig. \ref{fig:K1}(b) is $0$ for almost all values of $N$. This demonstrates that using extremely coarse 1-bit measurements does not prevent QCOMP from estimating the velocity of the target.
It is however worth noting that in this fairly simple setting where $K=1$, adding a dithering does not seem to help in terms of Estimation Error.
While this seems, at first glance, at odds with the theory developed in \cite{xu2018quantized} and mentioned in Section \ref{sec:quantize} it can be easily explained. The purpose of the developed method is to recover the frequency of the complex exponential, that is linked to the velocity of the only target, from 1-bit measurements. In a non dithered setting, this can be easily achieved by observing the rate at which the observed quadrants are changing. Adding a dithering, which allows us to have a bound on the reconstruction,  is still tantamount to lowering the SNR of the signal before quantization, which hinders this rate estimation process, for a given $M$. As it will become more apparent later, this explanation only holds for $K=1$.

\newlength{\wplot}
\setlength{\wplot}{0.5\columnwidth}
\newlength{\ngap}
\setlength{\ngap}{-0.0\columnwidth}
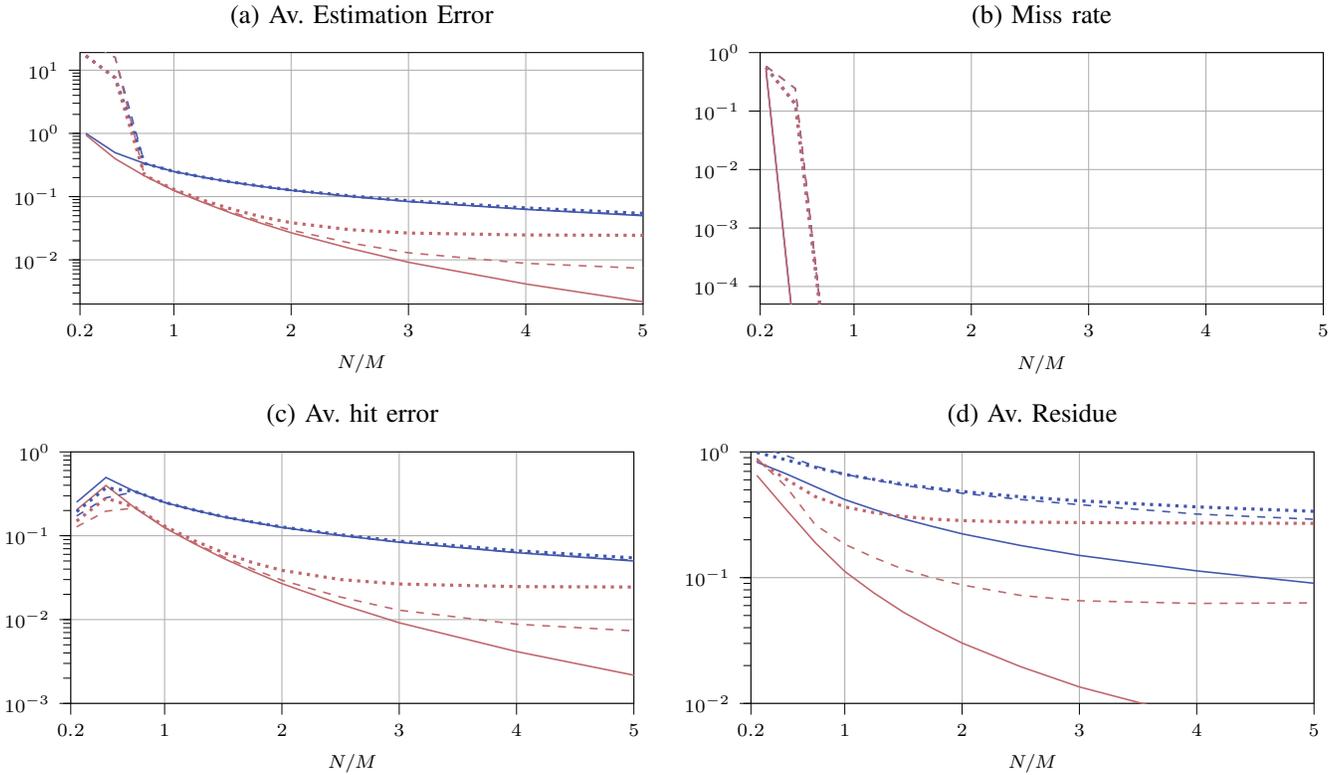
\begin{figure*}
\centering
\hspace{0.3\ngap}
\subfigure{\begin{tikzpicture}

\begin{axis}[
width=\wplot,
height=0.97*\htex,
tick align=outside,
tick pos=left,
log basis y={10},
xmajorgrids,
x grid style={white!69.01960784313725!black},
xlabel={\scriptsize $N/M$},
xmin=0.2, xmax=5,
xtick = {0.2,1,2,3,4,5},
xtick style={ color=black},
y grid style={white!69.01960784313725!black},
title={(a) Av. Estimation Error},
ymin=0.0020, ymax=19,
ymajorgrids,
ymode=log,
ytick style={color=black}
]

\addplot [semithick, bluen]
table {%
0.25 1.00007661608259
0.5 0.495196064906418
0.75 0.335132230845181
1 0.249416663002196
1.25 0.202357631457015
1.5 0.167737154157953
1.75 0.14436657131468
2 0.125608408491316
2.5 0.100647276733525
3 0.0836182785213524
4 0.0627576133453211
5 0.049939057862022
};
\addplot [semithick, redn]
table {%
0.25 0.945564210041207
0.5 0.397559961804968
0.75 0.21314185499157
1 0.124536087268236
1.25 0.0807205455696604
1.5 0.0538665952710391
1.75 0.0375523074414457
2 0.0268254431790585
2.5 0.0152129913341284
3 0.00915785436483331
4 0.00415469450300244
5 0.00216992902357482
};
\addplot [semithick, bluen, dashed]
table {%
0.25 29.6514669264814
0.5 15.903177789723
0.75 0.335155997666965
1 0.249503668464539
1.25 0.202493189992256
1.5 0.16829634018423
1.75 0.144959153677464
2 0.126146378810284
2.5 0.101193757170818
3 0.0841749026540035
4 0.0632977569062425
5 0.0509801938716191
};
\addplot [semithick, redn, dashed]
table {%
0.25 29.5964761744597
0.5 15.8137689949377
0.75 0.214554700578371
1 0.126030362655044
1.25 0.0825955198070279
1.5 0.056331915275133
1.75 0.0402238706241965
2 0.0294753401794019
2.5 0.0185475611210975
3 0.0129376885181851
4 0.00881113350559523
5 0.00734220503402566
};
\addplot [very thick, bluen, dotted]
table {%
0.25 16.9080579733182
0.5 7.57780813840652
0.75 0.335796145714908
1 0.250382424908421
1.25 0.203445971238886
1.5 0.16922046040125
1.75 0.145932938132192
2 0.127279775613655
2.5 0.102785302971663
3 0.0862721513235061
4 0.0662650440410766
5 0.0543481665830753
};
\addplot [very thick, redn, dotted]
table {%
0.25 16.8645652468377
0.5 7.48665895785153
0.75 0.217517098258122
1 0.130369784459474
1.25 0.0876762747452304
1.5 0.0629889360469378
1.75 0.0477929229642937
2 0.0386861831396846
2.5 0.0300097948391001
3 0.0265558291394836
4 0.0247466060768398
5 0.0244356576243006
};
\end{axis}

\end{tikzpicture}}
\hspace{\ngap}
\subfigure{\begin{tikzpicture}

\begin{axis}[
width=\wplot,
height=0.97*\htex,
tick align=outside,
tick pos=left,
log basis y={10},
xmajorgrids,
x grid style={white!69.01960784313725!black},
xlabel={\scriptsize $N/M$},
xmin=0.2, xmax=5,
xtick = {0.2,1,2,3,4,5},
xtick style={ color=black},
ytick = {0.00001,0.0001,0.001,0.01,0.1,1},
y grid style={white!69.01960784313725!black},
title={(b) Miss rate},
ymin=0.00005, ymax=1,
ymajorgrids,
ymode=log,
ytick style={ color=black},
legend style={nodes={scale=0.7, transform shape}}
]

\addplot [semithick, bluen]
table {%
0.25 0.498799999999961
0.5 0.00001
0.75 0
1 0
1.25 0
1.5 0
1.75 0
2 0
2.5 0
3 0
4 0
5 0
};
\addplot [semithick, redn]
table {%
0.25 0.498799999999961
0.5 0.00001
0.75 0
1 0
1.25 0
1.5 0
1.75 0
2 0
2.5 0
3 0
4 0
5 0
};
\addplot [semithick, bluen, dashed]
table {%
0.25 0.585099999999952
0.5 0.24259999999999
0.75 0.00001
1 0
1.25 0
1.5 0
1.75 0
2 0
2.5 0
3 0
4 0
5 0
};
\addplot [semithick, redn, dashed]
table {%
0.25 0.585099999999952
0.5 0.24259999999999
0.75 0.00001
1 0
1.25 0
1.5 0
1.75 0
2 0
2.5 0
3 0
4 0
5 0
};
\addplot [very thick, bluen, dotted]
table {%
0.25 0.559799999999955
0.5 0.132100000000002
0.75 0.00001
1 0
1.25 0
1.5 0
1.75 0
2 0
2.5 0
3 0
4 0
5 0
};
\addplot [very thick, redn, dotted]
table {%
0.25 0.558699999999955
0.5 0.131800000000002
0.75 0.00001
1 0
1.25 0
1.5 0
1.75 0
2 0
2.5 0
3 0
4 0
5 0
};
\end{axis}

\end{tikzpicture}}
\hspace{\ngap}
\subfigure{\begin{tikzpicture}

\begin{axis}[
width=\wplot,
height=0.97*\htex,
tick align=outside,
tick pos=left,
log basis y={10},
xmajorgrids,
x grid style={white!69.01960784313725!black},
xlabel={\scriptsize $N/M$},
xmin=0.2, xmax=5,
xtick = {0.2,1,2,3,4,5},
xtick style={ color=black},
y grid style={white!69.01960784313725!black},
title={(c) Av. hit error},
ymin=0.001, ymax=1,
ymajorgrids,
ymode=log,
ytick style={ color=black}
]

\addplot [semithick, bluen]
table {%
0.25 0.251074315673282
0.5 0.495196064906418
0.75 0.335132230845181
1 0.249416663002196
1.25 0.202357631457015
1.5 0.167737154157953
1.75 0.14436657131468
2 0.125608408491316
2.5 0.100647276733525
3 0.0836182785213524
4 0.0627576133453211
5 0.049939057862022
};
\addplot [semithick, redn]
table {%
0.25 0.203399666751293
0.5 0.397559961804968
0.75 0.21314185499157
1 0.124536087268236
1.25 0.0807205455696604
1.5 0.0538665952710391
1.75 0.0375523074414457
2 0.0268254431790585
2.5 0.0152129913341284
3 0.00915785436483331
4 0.00415469450300244
5 0.00216992902357482
};
\addplot [semithick, bluen, dashed]
table {%
0.25 0.172020885212066
0.5 0.283607369522252
0.75 0.335155997666965
1 0.249503668464539
1.25 0.202493189992256
1.5 0.16829634018423
1.75 0.144959153677464
2 0.126146378810284
2.5 0.101193757170818
3 0.0841749026540035
4 0.0632977569062425
5 0.0509801938716191
};
\addplot [semithick, redn, dashed]
table {%
0.25 0.127650593160291
0.5 0.196223053327173
0.75 0.214554700578371
1 0.126030362655044
1.25 0.0825955198070279
1.5 0.056331915275133
1.75 0.0402238706241965
2 0.0294753401794019
2.5 0.0185475611210975
3 0.0129376885181851
4 0.00881113350559523
5 0.00734220503402566
};
\addplot [very thick, bluen, dotted]
table {%
0.25 0.193980791648695
0.5 0.373314159257634
0.75 0.335796145714908
1 0.250382424908421
1.25 0.203445971238886
1.5 0.16922046040125
1.75 0.145932938132192
2 0.127279775613655
2.5 0.102785302971663
3 0.0862721513235061
4 0.0662650440410766
5 0.0543481665830753
};
\addplot [very thick, redn, dotted]
table {%
0.25 0.14944993963695
0.5 0.281423916448377
0.75 0.217517098258122
1 0.130369784459474
1.25 0.0876762747452304
1.5 0.0629889360469378
1.75 0.0477929229642937
2 0.0386861831396846
2.5 0.0300097948391001
3 0.0265558291394836
4 0.0247466060768398
5 0.0244356576243006
};
\end{axis}

\end{tikzpicture}}
\hspace{\ngap}
\subfigure{\begin{tikzpicture}

\begin{axis}[
width=\wplot,
height=0.97*\htex,
tick align=outside,
tick pos=left,
log basis y={10},
xmajorgrids,
x grid style={white!69.01960784313725!black},
xlabel={\scriptsize $N/M$},
xmin=0.2, xmax=5,
xtick = {0.2,1,2,3,4,5},
xtick style={ color=black},
y grid style={white!69.01960784313725!black},
title={(d) Av. Residue},
ymin=0.01, ymax=1,
ymajorgrids,
ymode=log,
ytick style={ color=black},
]

\addplot [semithick, bluen]
table {%
0.25 0.829836139304168
0.5 0.669821861974403
0.75 0.528261963510307
1 0.417644773522796
1.25 0.348412350909778
1.5 0.293325050261481
1.75 0.254927725704375
2 0.223199287926819
2.5 0.18014868512614
3 0.150296459343389
4 0.113240866203175
5 0.0902800670963042
};
\addplot [semithick, redn]
table {%
0.25 0.651590362998464
0.5 0.348349055504569
0.75 0.189036683015182
1 0.112115496074506
1.25 0.0754151581147776
1.5 0.0530875923386629
1.75 0.0394839317545001
2 0.0301950627788922
2.5 0.0195826373809414
3 0.0135188305226562
4 0.00769242093073652
5 0.00489193207896169
};
\addplot [semithick, bluen, dashed]
table {%
0.25 1.11707770751743
0.5 0.934949439421852
0.75 0.76933829205344
1 0.663791770007543
1.25 0.597737407454703
1.5 0.541735439937462
1.75 0.503521385406924
2 0.469458020607669
2.5 0.417914329549738
3 0.380605272960427
4 0.32039313551745
5 0.291275884039376
};
\addplot [semithick, redn, dashed]
table {%
0.25 0.890228192931293
0.5 0.528802628614323
0.75 0.262123551681154
1 0.184350837497757
1.25 0.144523614753212
1.5 0.116430473162234
1.75 0.0995141350184313
2 0.0875174897658086
2.5 0.07225062544239
3 0.0655107274638191
4 0.0623766089832905
5 0.0629796218805683
};
\addplot [very thick, bluen, dotted]
table {%
0.25 0.991132263659151
0.5 0.864900906186279
0.75 0.753115694347606
1 0.662839475521198
1.25 0.604539424390562
1.5 0.553324119533064
1.75 0.516735232843024
2 0.484910229910571
2.5 0.439601554120152
3 0.408535191340475
4 0.365467791308598
5 0.336891128771623
};
\addplot [very thick, redn, dotted]
table {%
0.25 0.858439241090964
0.5 0.60630051450275
0.75 0.440554728458611
1 0.365147417601485
1.25 0.32872379027291
1.5 0.306521462387118
1.75 0.291812339347085
2 0.284378988618225
2.5 0.27637736913032
3 0.274231273184796
4 0.271896336526267
5 0.270264715405616
};
\end{axis}

\end{tikzpicture}}
\hfill
\caption{Performance metrics comparison between the application of QCOMP with non-interpolated (\textcolor{bluen}{blue}) and the interpolated (\textcolor{redn}{red}) scheme working either with the full resolution measurement $\bs y$ (solid), the non-dithered quantized measurement (dashed) or the dithered quantized measurement (dotted) for $K=1$. \textbf{(a)} Average error on the velocity estimation, \textbf{(b)} miss rate, \textbf{(c)} average hit error and  \textbf{(d)} average value of the residual reconstruction error.}
\label{fig:K1}
\end{figure*}

\begin{figure*}
\centering
\hspace{0.3\ngap}
\subfigure{\begin{tikzpicture}

\begin{axis}[
width=\wplot,
height=0.97*\htex,
tick align=outside,
tick pos=left,
log basis y={10},
xmajorgrids,
x grid style={white!69.01960784313725!black},
xlabel={\scriptsize $N/M$},
xmin=0.2, xmax=5,
xtick = {0.2,1,2,3,4,5},
xtick style={ color=black},
y grid style={white!69.01960784313725!black},
title={(a) Av. Estimation Error},
ymin=0.0020, ymax=19,
ymajorgrids,
ymode=log,
ytick style={color=black}
]

\addplot [semithick, bluen]
table {%
0.25 1.45664044998238
0.5 0.66556869210024
0.75 0.450908532368672
1 0.465943177686397
1.25 0.406132937385712
1.5 0.30954184487011
1.75 0.247211849901494
2 0.196687438132425
2.5 0.162673965117519
3 0.131302347853746
4 0.0929716064304517
5 0.07750847154193
};
\addplot [semithick, redn]
table {%
0.25 1.44357951923013
0.5 0.504159405059288
0.75 0.317350398132413
1 0.163574545093435
1.25 0.100871524302462
1.5 0.0634506881351738
1.75 0.0514330896580102
2 0.0337195711734603
2.5 0.0208316251739555
3 0.0146652461472433
4 0.00877434394305617
5 0.0069051745238374
};
\addplot [semithick, bluen, dashed]
table {%
0.25 20.8800470131031
0.5 12.2107693224013
0.75 6.06818787540107
1 5.29105580766653
1.25 5.39489820687771
1.5 5.33672829261725
1.75 5.14952117344514
2 5.36685119676474
2.5 5.28921410177697
3 5.15302649898868
4 5.0764537391976
5 4.77235463815222
};
\addplot [semithick, redn, dashed]
table {%
0.25 20.6505346967754
0.5 12.8321069574354
0.75 5.88194226912496
1 2.34644664810647
1.25 2.7260263998664
1.5 2.83009085950681
1.75 3.40984969267424
2 3.73151239735685
2.5 4.36672556483242
3 4.66818533192494
4 4.96070171513953
5 4.78133414390692
};
\addplot [very thick, bluen, dotted]
table {%
0.25 28.0307260376586
0.5 12.5362815708328
0.75 1.32235916992901
1 0.635538229224671
1.25 0.645043743104926
1.5 0.559708940583897
1.75 0.548902324240639
2 0.568593622895274
2.5 0.571161094421803
3 0.442183526989969
4 0.406848691239214
5 0.396405816075049
};
\addplot [very thick, redn, dotted]
table {%
0.25 24.5854053271627
0.5 12.8944102070807
0.75 1.34752096495953
1 0.499529826244419
1.25 0.467224321343297
1.5 0.354182194342842
1.75 0.323024439173911
2 0.270948154047531
2.5 0.255265411553325
3 0.210940008094922
4 0.211655439039551
5 0.203240060953255
};
\end{axis}

\end{tikzpicture}}
\hspace{\ngap}
\subfigure{\begin{tikzpicture}

\begin{axis}[
width=\wplot,
height=0.97*\htex,
tick align=outside,
tick pos=left,
log basis y={10},
xmajorgrids,
x grid style={white!69.01960784313725!black},
xlabel={\scriptsize $N/M$},
xmin=0.2, xmax=5,
xtick = {0.2,1,2,3,4,5},
ytick = {0.00001,0.0001,0.001,0.01,0.1,1},
xtick style={ color=black},
y grid style={white!69.01960784313725!black},
title={(b) Miss rate},
ymin=0.00005, ymax=1,
ymajorgrids,
ymode=log,
ytick style={ color=black}
]

\addplot [semithick, bluen]
table {%
0.25 0.525749999999951
0.5 0.118749999999996
0.75 0.0621000000000013
1 0.128699999999994
1.25 0.105999999999996
1.5 0.0782499999999995
1.75 0.0504000000000009
2 0.0357500000000005
2.5 0.0290500000000003
3 0.0205000000000001
4 0.0106
5 0.00714999999999998
};
\addplot [semithick, redn]
table {%
0.25 0.524249999999951
0.5 0.0530000000000009
0.75 0.0397500000000006
1 0.01685
1.25 0.00355
1.5 0.00205
1.75 0.00165
2 0.00065
2.5 0.0002
3 0.00015
4 0
5 5e-05
};
\addplot [semithick, bluen, dashed]
table {%
0.25 0.630599999999942
0.5 0.36624999999997
0.75 0.292349999999976
1 0.265999999999979
1.25 0.270349999999978
1.5 0.227799999999983
1.75 0.213349999999985
2 0.192199999999987
2.5 0.182399999999988
3 0.176699999999989
4 0.171199999999989
5 0.170799999999989
};
\addplot [semithick, redn, dashed]
table {%
0.25 0.631799999999942
0.5 0.343099999999973
0.75 0.25379999999998
1 0.208049999999985
1.25 0.186099999999988
1.5 0.179199999999988
1.75 0.175049999999989
2 0.175899999999989
2.5 0.173949999999989
3 0.172199999999989
4 0.16874999999999
5 0.169899999999989
};
\addplot [very thick, bluen, dotted]
table {%
0.25 0.631449999999942
0.5 0.281199999999982
0.75 0.091999999999998
1 0.128749999999994
1.25 0.110499999999996
1.5 0.0866999999999986
1.75 0.0612000000000012
2 0.051250000000001
2.5 0.0373500000000006
3 0.0288500000000003
4 0.0222000000000001
5 0.01665
};
\addplot [very thick, redn, dotted]
table {%
0.25 0.65364999999994
0.5 0.239499999999988
0.75 0.064150000000001
1 0.0326500000000003
1.25 0.0223500000000001
1.5 0.0192
1.75 0.0152999999999999
2 0.01865
2.5 0.0179
3 0.0156499999999999
4 0.0142999999999999
5 0.01695
};
\end{axis}

\end{tikzpicture}}
\hspace{\ngap}
\subfigure{\begin{tikzpicture}

\begin{axis}[
width=\wplot,
height=0.97*\htex,
tick align=outside,
tick pos=left,
log basis y={10},
xmajorgrids,
x grid style={white!69.01960784313725!black},
xlabel={\scriptsize $N/M$},
xmin=0.2, xmax=5,
xtick = {0.2,1,2,3,4,5},
xtick style={ color=black},
y grid style={white!69.01960784313725!black},
title={(c) Av. hit error},
ymin=0.001, ymax=1,
ymajorgrids,
ymode=log,
ytick style={ color=black}
]

\addplot [semithick, bluen]
table {%
0.25 0.462527352031802
0.5 0.487622172769633
0.75 0.358416069145364
1 0.257324588673767
1.25 0.224071126513697
1.5 0.170631013180605
1.75 0.155798971447942
2 0.131134122816257
2.5 0.106323163525917
3 0.0911218951490191
4 0.0723325084444019
5 0.0623352894893872
};
\addplot [semithick, redn]
table {%
0.25 0.391732744565571
0.5 0.390287672793538
0.75 0.252821213459567
1 0.134749833211599
1.25 0.0945999062857146
1.5 0.0598688403446283
1.75 0.0479831866702172
2 0.0324362803820288
2.5 0.0204825558078285
3 0.0143677841211897
4 0.00877434394305617
5 0.00656741463223078
};
\addplot [semithick, bluen, dashed]
table {%
0.25 0.309864346611308
0.5 0.386223666853899
0.75 0.328018754099443
1 0.260380078069153
1.25 0.230871975506353
1.5 0.178124019940048
1.75 0.165041429035095
2 0.141241690419887
2.5 0.119006207458125
3 0.104570393340138
4 0.0878175469523636
5 0.0804262982149792
};
\addplot [semithick, redn, dashed]
table {%
0.25 0.259786898124719
0.5 0.302214976358665
0.75 0.22236925941765
1 0.141463640235967
1.25 0.102935280419747
1.5 0.0748338922603808
1.75 0.0634883559561042
2 0.0539010574053767
2.5 0.046376866674774
3 0.0428629827503858
4 0.0401195927848196
5 0.0398639837567397
};
\addplot [very thick, bluen, dotted]
table {%
0.25 0.291029922880576
0.5 0.343300086128502
0.75 0.348001377468731
1 0.260782429492741
1.25 0.229776218558237
1.5 0.176142767480565
1.75 0.164102754875519
2 0.137404545917874
2.5 0.115547937359704
3 0.101524260282427
4 0.0856046407081398
5 0.0783847250350905
};
\addplot [very thick, redn, dotted]
table {%
0.25 0.22078889336021
0.5 0.259561446800705
0.75 0.253884958423435
1 0.160402454002139
1.25 0.11934581162646
1.5 0.0973614536236417
1.75 0.0860036598306733
2 0.0780602385326858
2.5 0.0714586608198141
3 0.0688312523905735
4 0.0663100449073003
5 0.0657141969922646
};
\end{axis}

\end{tikzpicture}}
\hspace{\ngap}
\subfigure{\begin{tikzpicture}

\begin{axis}[
width=\wplot,
height=0.97*\htex,
tick align=outside,
tick pos=left,
log basis y={10},
xmajorgrids,
x grid style={white!69.01960784313725!black},
xlabel={\scriptsize $N/M$},
xmin=0.2, xmax=5,
xtick = {0.2,1,2,3,4,5},
xtick style={color=black},
y grid style={white!69.01960784313725!black},
title={(d) Av. Residue},
ymin=0.01, ymax=1,
ymajorgrids,
ymode=log,
ytick style={ color=black},
]

\addplot [semithick, bluen]
table {%
0.25 0.853370749888327
0.5 0.658629589105202
0.75 0.563829303296266
1 0.373540750463809
1.25 0.352487702085433
1.5 0.257112751118602
1.75 0.26376122618389
2 0.20494877876757
2.5 0.175038988453862
3 0.152746975314707
4 0.126300110431767
5 0.112233972446189
};
\addplot [semithick, redn]
table {%
0.25 0.703836831561763
0.5 0.286261613824375
0.75 0.163618097822081
1 0.0874097171984528
1.25 0.0758050462674196
1.5 0.0438098689162984
1.75 0.0425471777520472
2 0.0271454231051778
2.5 0.0194609775037146
3 0.0151179895992922
4 0.0106240566547056
5 0.00859905588165626
};
\addplot [semithick, bluen, dashed]
table {%
0.25 1.14231832499447
0.5 0.969920128214049
0.75 0.835623016162862
1 0.622497118369024
1.25 0.606823014873166
1.5 0.549765949707577
1.75 0.558396902717542
2 0.524195185429875
2.5 0.512081509182952
3 0.501474455252865
4 0.485890898434804
5 0.477842739622132
};
\addplot [semithick, redn, dashed]
table {%
0.25 0.958112762279471
0.5 0.604823104270485
0.75 0.483100270636531
1 0.45132208147846
1.25 0.441388952622266
1.5 0.437879940975044
1.75 0.4349042852771
2 0.433317232551837
2.5 0.433871334706401
3 0.433721936692469
4 0.433257301282852
5 0.433050332142416
};
\addplot [very thick, bluen, dotted]
table {%
0.25 0.87386497466808
0.5 0.743792989490012
0.75 0.692473736636982
1 0.559326932719854
1.25 0.55409950483544
1.5 0.481710264429406
1.75 0.48812707429166
2 0.444634074903975
2.5 0.423119970754897
3 0.408547869773872
4 0.388401735957307
5 0.381293686849606
};
\addplot [very thick, redn, dotted]
table {%
0.25 0.817673004802529
0.5 0.560204663259112
0.75 0.434679316877437
1 0.393772972257324
1.25 0.383863666215971
1.5 0.374368826741672
1.75 0.370617928782165
2 0.367823185094259
2.5 0.36669924664445
3 0.364657719462392
4 0.362720346553786
5 0.364756242874286
};
\end{axis}

\end{tikzpicture}}
\hfill
\caption{Performance metrics comparison between the application of QCOMP with non-interpolated (\textcolor{bluen}{blue}) and the interpolated (\textcolor{redn}{red}) scheme working either with the full resolution measurement $\bs y$ (solid), the non-dithered quantized measurement (dashed) or the dithered quantized measurement (dotted) for $K=2$. \textbf{(a)} Average error on the velocity estimation, \textbf{(b)} miss rate, \textbf{(c)} average hit error and  \textbf{(d)} average value of the residual reconstruction error.}
\label{fig:K2}
\end{figure*}
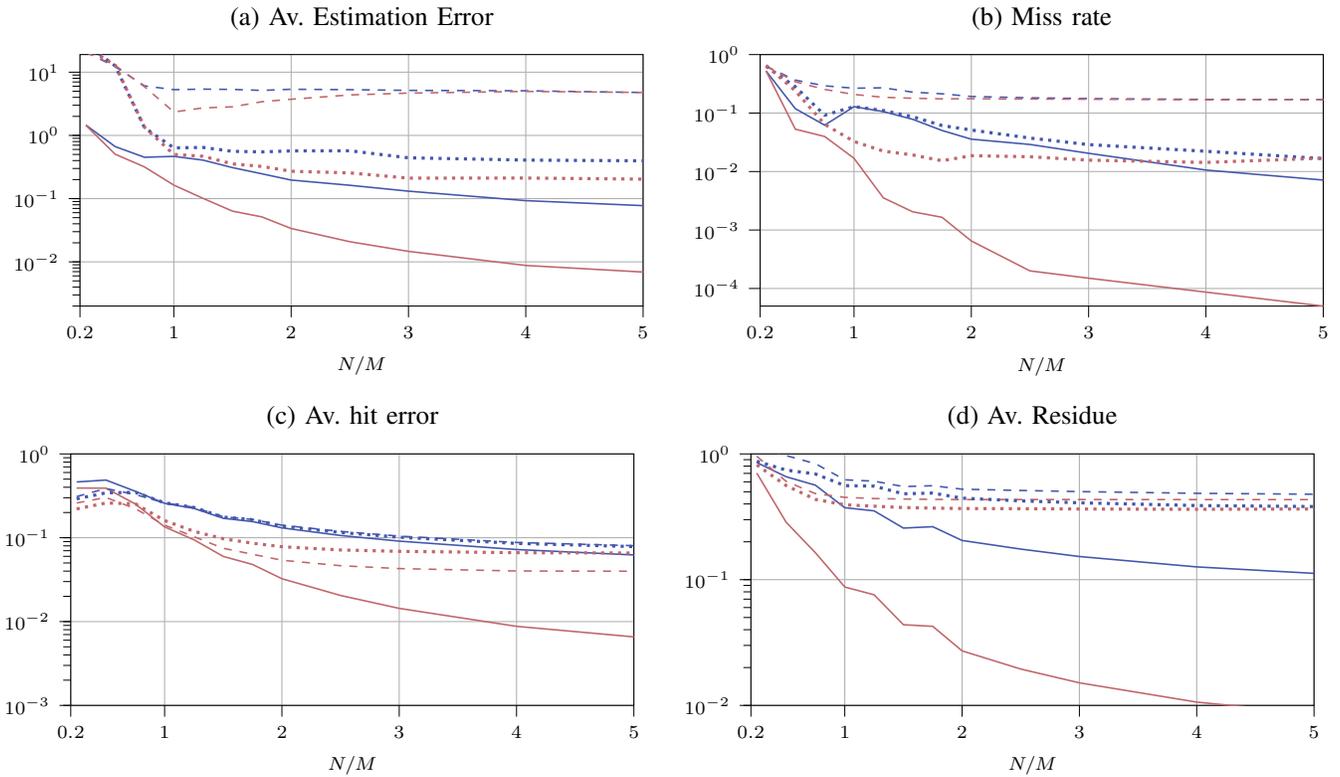

Increasing the number of targets to $K=2$ in Fig.\ref{fig:K2} shows the real potential of the \textit{Off-the-Grid} estimation from quantized and dithered targets. Indeed, in Fig.\ref{fig:K2}(d) we see that the tendencies observed in Fig.\ref{fig:K1}(d), are reversed, \ie adding a dithering to the quantization enhances the reconstruction compared to the deterministic quantization.
In Fig.\ref{fig:K2}(a) we see, for both interpolation schemes, that the dithered scheme outperforms the determistic quantization by an order of magnitude. In Fig.\ref{fig:K2}(b), we see that the deterministic scheme saturates at miss rate of around $20\%$, while the dithered scheme approaches the $1\%$. This is consistent with the theory developed in \cite{xu2018quantized,Feuillen2018QuantityOQ} and referenced in Section \ref{sec:quantize}, where it is shown that a coarse and deterministic quantization induces artefact in the quantization that often translates to cancelling the effect of one or more target from the signal.
In other terms, the miss rate is higher without dithering because in many realisations, QCOMP only recovers one of the targets.
In Fig.\ref{fig:K2}(c), we see that adding a dither ensures that most of the targets will be recovered, but at a price in the quality of the estimation.

We finish this section by noting that in Fig.\ref{fig:K2}(a-d) the 1-bit schemes seems to saturate when the grid density $N/M$ increases for a fixed $M=256$, while no saturation is observed with the full resolution scheme.
Increasing $N/M$ improves the quality of the approximation \eqref{eq:interp_scheme} and hence, reduces the interpolation error which leads to a lower miss rate.
However the estimation error induced by the quantization are better mitigated by the dithering for a large value of $M$. 
Therefore, when $M$ is constant, this \emph{quantization} error remains constant and begins to dominate the \emph{interpolation} error when $N/M$ increases.

\section{Conclusion and future works}
\label{sec:concl}

In this paper we combined two distinct fields in radar detection theory: one where we dramatically lower the resolution of the ADCs to obtain a 1-bit quantizer and one where we look for target parameters that exist in a continuous space.
To this end, we extended the existing Continuous OMP to take into account such 1-bit quantizer by combining the dithering and interpolation strategies. 
We proposed the Quantized Continuous OMP (QCOMP) in its simplest and fastest definition.
We may investigate in the future more sophisticated definitions of the QCOMP that propose a better trade-off between accuracy and computation time.
Yet our simple definition of QCOMP already shows that the estimation of velocities from a continuous domain exhibits behaviours with respect to the dithering that are similar to those previously studied in the \ong framework where the real target velocities lie in a discrete space.
Moreover, we observed and explained how working with the quantized measurements modifies the sensitivity of the algorithm to the grid density when velocities lie \ofg. 
In the future, we will extend this work to more complex radar systems allowing the estimation of ranges, velocities and angles. 

\section*{Acknowledgment}
The authors would like to thank the COSY meeting in ICTEAM that gave us the opportunity to discuss and find the idea for this paper. 

\bibliographystyle{unsrt}
\begin{small}
\bibliography{biblio.bib,QCS_SOTA.bib}

\end{small}

\end{document}